\documentstyle[eqsecnum,aps]{revtex}

\begin{document}
\draft
\preprint{HEP/123-qed}
\title{
Fractal versus QuasiClassical Diffusive Transport
\\
in a class of quantum systems
}
\author{Fausto Borgonovi\\}
\address{
Dipartimento di Matematica, Universit\`a Cattolica, sede di Brescia,
via Trieste 17, 25121 Brescia, Italy  \\
Istituto Nazionale di Fisica Nucleare, Sezione di Pavia,
via Bassi 6, 27100 Pavia, Italy.
}
\author{Italo Guarneri}
\address{
Universit\`a di Milano, sede di Como, via Lucini 3, 22100 Como, Italy \\
Istituto Nazionale di Fisica Nucleare, Sezione di Pavia,
via Bassi 6, 27100 Pavia, Italy.
}
\date{\today}
\maketitle
\begin{abstract}
 We compare the properties of transmission across one-dimensional
finite samples which are associated with  two types of "quantum diffusion", one
related to a classical chaotic dynamics, the other to a multifractal
energy spectrum. We numerically investigate models exhibiting one or both
of these features, and we find in all cases an inverse power law dependence
of the average
transmission on the sample length. Although in all the considered cases
the quadratic spread of wave packets increases linearly (or very close
to linearly) in time for both types of dynamics, a proper Ohmic dependence is
always observed only in the case of quasi-classical diffusion. The analysis
of the statistics
 of transmission fluctuations in the case of a fractal spectrum exposes some
new features, which mark further differences from ordinary diffusion, and
 enforce the conclusion that the
two types of transmission are intrinsically different.
\end{abstract}


\twocolumn

\section{Introduction}
\label{sec:level1}
The onset of chaos in classical single-particle dynamics is often accompanied
by deterministic diffusion, which is  a type of
stochastic motion which mimics to some extent a random walk, even in the
absence of
external random agents. Typical examples
are the motion of a particle elastically bouncing in an array
of fixed
scatterers, and the diffusive energy absorption occurring in nonlinear
periodically driven systems. In both cases, characteristic transport
properties appear, that can be more or less accurately described by an
equation of the Fokker-Planck type. Whether similar transport properties
persist on the quantum level is an important question, which lies on the
borderline between Quantum Chaology and Solid State physics, and has received
much attention on both sides.
A most remarkable result in this context is that chaotic diffusion is inhibited
by quantization. A classical one-dimensional diffusion
is quantally reproduced
only on a finite time scale, and is in the long run stopped
by an interference effect very similar to Anderson localization.
 Equivalently, if the transmission of
particles across a finite sample of a quasi-one-dimensional
 disordered solid is
considered,  then the inverse dependence on the sample length characteristic of
a classical diffusive transport
(Ohm's law) is quantally observed only on length scales
which are
small in comparison with the Anderson localization length.

It was realized that, in a quasi-one dimensional case,  an unbounded
 diffusive (or anomalously diffusive) spreading
of wave packets is often associated with a singular continuous
spectrum of the Hamiltonian (or the Floquet operator
in the case of periodically driven systems).
This sort of a spectrum, regarded as a mathematical curiosity a
long time, appears quite frequently in incommensurate structures, such as
involved e.g. by quasi-crystals and by electron dynamics in crystals in
magnetic fields; in such cases fractal singular spectra
of zero Lebesgue measure have been found.
In the light of the above findings, and insofar as the quasi-one
dimensional
case is concerned, a quantum "diffusive" regime, i.e., one marked by a linear
(or close to linear) increase in time of the mean square displacement, can
exist
either due to an underlying classical chaotic diffusion, developing on small
scales in which localization effects are not yet apparent, or due to a fractal
structure of the spectrum; no scale limitations being necessary in the latter
case.

Prototype systems for these two situations are the Kicked Rotator
(KRM) and the
Harper (HM)
 model, respectively. Each of these two models exhibits one type of
quantum diffusion, but not the other; therefore, the two phenomena are in
principle independent of each other. Still,
the question remains, whether the "fractal"
unbounded diffusion may offer a quantum counterpart for the classical unbounded
propagation in cases when both are present. A particular aspect of this
question is what kind of transmission properties are associated with the
"fractal" diffusion, and how do they match with the ohmic transport typical
of classical diffusion.

We have studied this question using the two mentioned models plus a third one,
which in a sense interpolates between them. This is the
Kicked Harper Model (KHM),
which displays at once a chaotic classical limit marked by deterministic
diffusion (which is missing in the HM), and a "critical" quantum
regime with a multifractal quasi-energy
spectrum (which is missing in the KRM).
We have compared the properties of quantum transmission across finite samples
which are observed in the three cases.
The analogies and differences we have observed  supply useful
indications for answering the questions raised before.
For example, according to our results, the Ohm law seems to
be rather exceptional inside the class of quantum systems
endowed with "fractal" quantum diffusion. The fingerprint of this
macroscopic law would be then represented by the concurrent
presence of a quasi-classical diffusion.

Other differences emerge on analyzing transmission fluctuations, and lead to
the conclusion that the two types of diffusion are marked by essentially
different statistical properties, and are in no apparent relation
to each other.

The plan of the paper is the following:
in Sec.\ \ref{sec:level2} we review the mathematical formalism
which allows for a general treatment of the kicked dynamics
as a scattering process.
In  Sec.\ \ref{sec:level3}
we present the old (A) and the new (B,C) results about the transport
properties respectively in the KRM, in the KHM and in the HM.
The conclusive sec. IV is devoted to the analysis of
the transmission fluctuations.
\vskip 1cm

\section{Discrete-time scattering processes}
\label{sec:level2}

We are interested in one-dimensional scattering processes in which  free
particles impinge on a "sample" of a finite length, whereby they are either
reflected
or transmitted. The particle dynamics inside the sample will have a nontrivial
diffusive character, both in the classical and in the quantum case.
 Such a
situation cannot of course occur in  strictly one-dimensional, conservative
Hamiltonian cases,  but it
can be realized with relative ease if one considers a discrete-time dynamics
instead of a continuous-time one - an artifact which has proven extremely
useful in Nonlinear Dynamics.
We shall presently outline  the general formalism needed to describe a
discrete-time scattering process.

We consider the discrete time dynamics of a particle on a line, with
coordinate $q$ and momentum $p$. Throughout this paper we shall use
dimensionless variables. The classical dynamics of the particle is described
by the iteration of an area-preserving map, of the form:
\begin{equation}
\begin{array}{lcl}
\label{map}
\bar{ q} &= &q-{{\partial S(p)}\over{\partial p}}\\
\bar{ p} &= &\left\{
\begin{array} {ll}
       p+{{\partial V(\bar q)}\over{\partial \bar q}}&
       \mbox{if $ q_0 \leq {\bar q}\leq q_0+L$}\\
       p&
       \mbox{elsewhere}
\end{array}
\right.
\end{array}
\end{equation}
where the generating functions $S,V$ will have different forms,
depending on the considered model. In all cases, however,
the function $S$ will be even and
periodic in the momentum $p$, with period $2\pi$. Outside the region $
\quad q_0
\leq q\leq q_0+L$ the motion is uniform at constant speed
$-S'(p)$ and
momentum $p$. Thus
eqs. (\ref{map}) define a scattering problem, in which free particles coming
from infinity impinge on the scattering region, wherein they undergo
"collisions" described by  the "potential" $V(q)$, until they re-exit from
either end.
With appropriate
choices of the generating functions $S,V$, the dynamics inside the sample
will have a chaotic character, leading to fast randomization in $p$ and
to diffusion in $q$.

 In this case, the scattering
trajectories sensitively depend
on various parameters, including $q_0$, which defines the position of the
scatterer on the line; both transmission and reflection are
affected in an unpredictable way by small changes of $q_0$,
 because $q_0$, together with the incoming momentum,
 defines the initial conditions for the
chaotic map
whose iteration defines the dynamics of the particle inside the scatterer.

For this reason, we can generate statistical ensembles by placing the
scatterer at different positions; different locations of the scatterer
 of length $L$ will define different statistical samples
 in the ensemble. This is in clear analogy
with the problem
of electronic transmission through disordered solids. In that case
 different
samples correspond to different realizations of disorder.

The Quantization of the dynamics (\ref{map}) is obtained in a standard way
by iterating the quantum one-step
propagator:
\begin{equation}
\label{U}
\hat{U} = e^{{i\over\hbar}V({\hat q})} e^{{i\over\hbar}S ({\hat p})}
\end{equation}
where we have introduced the adimensional Planck constant $\hbar$, and
$V(q)$ is taken $\not=0$ only for $q$ inside the scattering region.

Due to periodicity in $p$, the Bloch theorem applies, and the evolution
(\ref{U}) preserves the "quasi-position" ${\hat\alpha}={\hat q}$
The value  $\alpha$ of the quasi-position is a constant of the  motion,
therefore its value can be arbitrarily fixed. The quantum dynamics at
fixed $\alpha$ can be studied, by using the quantization rule:
\begin{equation}
\hat q_{\alpha}=-i\hbar{{\partial}\over{\partial p}}+\alpha
\end{equation}
defined on $2\pi$-periodic wave functions.
Quantized in this way, the position  assumes discrete values $n\hbar+\alpha$,
 ($n$ integer), that is, the quantum motion at given quasi-position takes place
on a discrete one-dimensional lattice. We shall denote $\vert n\rangle$ the
eigenstate corresponding to the $n-$th site of this lattice. The quantum
 evolution
at fixed $\alpha$ is given by a unitary operator ${\hat U}_\alpha$ obtained
by writing ${\hat q}_\alpha$ instead of ${\hat q}$ in eqn.(\ref{U}). Whereas
${\hat U}$ acts in the space of square-summable functions on the line,
${\hat U}_\alpha$ acts in the space of square-summable functions on the
lattice.

The unitary operator ${\hat U}_\alpha$ has eigenvalues
$e^{i\lambda/\hbar}$, and
$\lambda\quad(mod\quad 2\pi\hbar)$ is known as the
quasi-energy. The quasi-energy is obviously conserved under the evolution.

The quantum scattering theory stems, as usual, from a comparison of the long-
time asymptotics of the dynamics generated by (\ref{U}) with the free
dynamics generated by ${\hat U_0}=\exp(iS({\hat p})/\hbar)$. On
mathematical grounds,  ${\hat U}_\alpha$ is a finite-rank perturbation of
${\hat U_0}$. This entails the existence of the M\o{}ller wave operators
${\hat \Omega}_{\mp}=\lim_{t\to\pm\infty}{\hat U}^t{\hat U}_0^{-t}$ and
thus ensures the
 well-posedness of the scattering problem.
 In order to find the Scattering Matrix it is
convenient to resort to time-independent scattering theory, and to seek the
scattering eigenfunctions of ${\hat U}$ corresponding to a given value
$\lambda$ of the quasi-energy.
The eigenfunctions of the free dynamics corresponding to the same quasi-
energy are:
\begin{equation}
\langle n \vert u_0^{\lambda,j} \rangle =
(2\pi)^{-1/2} e^{-i n p_j/\hbar}
\end{equation}
where $p_j,\ (j=1,...N_\lambda) $ are the roots of the equation
\[S(p_j) = \lambda \ (mod \ 2\pi\hbar). \]

In this way for each
quasi-energy $\lambda $ there are  $N_\lambda$
degenerate eigenfunctions; every one of them corresponds to an open scattering
channel at the given quasi-energy. The number $N_\lambda$ of such channels
depends on the quasi-energy $\lambda$ and on the specific form of the function
$S$.

The scattering dynamics can be visualized as follows~: plane waves,
eigenfunctions
of ${\hat U}_0$, are scattered by the sample, thus giving rise
to reflected and transmitted waves. The crucial feature of this process is that
quasi-energy is conserved, that is, $S(p)$ is changed by an integer multiple of
$2\pi\hbar$.
The construction of the ${\cal S}$-matrix follows standard rules
of the scattering theory, via the Lippmann--Schwinger
equation :
\begin{equation}
[1-e^{i\lambda/\hbar}{\hat G}_{+}(e^{-i{\hat V}/\hbar}-1)]u_{+}^{\lambda,j}
= u_{0}^{\lambda,j}\label{LiSc}
\end{equation}
relating free waves $u_0$ with true, distorted waves  $u_+$.
In the above formula $\hat{G}_{+}$ is the ``free'' Green
function :
\begin{equation}
{\hat G}_+ (\lambda) = \lim_{\epsilon \to {0^+} }
( e^{i{\hat S}/\hbar}-e^{i\lambda/\hbar+\epsilon})^{-1}
\end{equation}

By inverting the system  (\ref{LiSc}) one obtains the distorted waves $u_+$
and the ${\cal S}$--matrix, whose elements are given by~:
\begin{equation}
{\cal S}_{j,k} = \delta_{j,k} - 2\pi \vert \nu_j \vert^{1/2}
\vert \nu_k \vert^{1/2} \langle
(e^{i{\hat V}/\hbar}-1)u_{0}^{\lambda,j} \vert
u_{+}^{\lambda,k} \rangle
\end{equation}
where $j,k=1,\ldots N_\lambda$ and
$\vert \nu_j \vert = 1/ \vert dS(p_j)/dp \vert $
is the density of states.
Given the ${\cal S}$--matrix, the transmission coefficient is
obtained by summing the squared moduli of the $\cal S$--matrix elements
related to transmission in a given direction of propagation.

The ${\cal S}-$matrix is sample-dependent because the
explicit form of ${\hat U}_\alpha$ depends on the choice of the
sample, that is, on $q_0$, through the values of the function
$A(q)= \exp(i V(q)/\hbar)$
at $q=n\hbar+\alpha, \quad(q_0\leq q\leq q_0+L)$. The
string of these $\sim L/\hbar$ values has a more or less disordered character,
depending on the choice of the function $V(q)$; in any case, for any
nontrivial choice of the latter,
different samples correspond to different strings, and different strings
correspond to different realizations of "disorder".
There is a connection between the dependence of the ${\cal S}$--matrix on the
sample and its dependence on the quasi-position $\alpha$, because changing
the latter by $\hbar$ is equivalent to shifting the sample by one site along
the $q-$lattice.
This suggests that the above described process of
averaging over disorder may be equivalent to averaging over all possible
values of the quasi-position. This is certainly true when $2\pi/\hbar$ is
an irrational number and
$A(q)$ is a
$2\pi$-periodic function - a case that occupies a central position in this
paper.  In that case, any quantity related to scattering is
also a $2\pi-$periodic function of $\alpha$, and its average over all
samples at
given $\alpha=\alpha_0$ coincides with the average over all the values of
$\alpha=\alpha_0+n\hbar\quad
mod(2\pi)$, which are obtained from $\alpha_0$ by iterating
the shift $\alpha\to\alpha\hbar\quad mod(2\pi)$. Indeed, since this shift is
ergodic
for $2\pi/\hbar$ irrational, such "time-averages"  are independent of
the starting point $\alpha_0$ and
coincide with "phase-averages", i.e., averages over $\alpha_0$.

\section{Transport properties }
\label{sec:level3}

\subsection{The Kicked Rotator }

With the choice $S(p)=b\cos p$,
$V(q)={\tau\over 2} q^2$ the map (\ref{map})
describing the dynamics inside the sample becomes the Standard Map, which
has been widely studied both in its classical
and in its quantum version; the latter is associated with a well-known
 dynamical model,
known as the KRM.

The quantum KRM is a paradigm of the quantum
phenomena associated with classical
chaotic diffusion, and a short review of its main properties will
 provide a proper frame for the following discussion.
At $b\tau \gg 1$ the classical Standard Map is fully chaotic,
 and the $q-$motion is similar to a random walk,
with a
diffusion coefficient
$D=\beta(b)b^2/2$ where $\beta(b)$ is a
known bounded function of $b$\cite{LL}.
On length scales much bigger than $b$,
 the distribution $f(q,t)$ of
a statistical ensemble of orbits is adequately described by the diffusive
approximation, which assumes $f(q,t)$ to satisfy the diffusion
(Fokker-Planck) equation \cite{LL,CHI,RW}. \par

The properties of the quantum KRM
 (at fixed quasi-position) crucially
depend on the arithmetic nature of $\hbar$. If $\hbar$ is a rational multiple
 of $4\pi$, then the evolution operator (\ref{U}) has an absolutely
 continuous spectral component that produces a ballistic spread of
 wave-packets on the $q-$lattice. If $\hbar$ is a "good" irrational number,
accumulated
 numerical evidence supports a pure-point spectrum, and exponential
 localization of wave-packets on the lattice, with a localization length
 $\ell\sim b^2/\hbar$
(the properties of the KRM are reviewed in \cite{Felix}).
 A continuous spectral
 component is also known to exist when $\hbar/4\pi$ is a "bad" irrational,
 well approximated by rationals; however, neither the exact type of
 approximation required, nor the nature of the corresponding continuous
spectrum are
 known.
 Thus in all known cases the long-time nature of the quantum propagation is
 different from that of the classical motion; while the latter is diffusive,
 the former is either ballistic or localized. Since the
 Correspondence Principle
entails that in the limit $\hbar\to 0$ the classical
 diffusive motion must be recovered, there must be a time scale
 $t_D(\hbar)$ separating the quasi-classical diffusive spread of wave
 packets from the purely quantum regime, (ballistic or localized). It is
 known that $t_D\sim D/\hbar$, which is the time required to diffuse over
the localization length $\ell$.

The scattering problem associated with the KRM was
 investigated in detail in \cite{BOGU}.
In the classical case, for $b\ll L$ the diffusive
 approximation is valid, and the
numerically computed transmission coefficient $T$ is in excellent
 agreement with the theoretical value:
\begin{equation}
{1\over T } = 2-{1\over \beta}+{ {2b}\over \pi D }L\label{trans}
\end{equation}
 which is found by solving the diffusion equation in the sample of length
 $L$, subject to appropriate boundary conditions \cite{BOGU}.
 On comparing (\ref{trans}) with the kinetic formula $T\approx \ell_0/L$
 involving the mean free path (mfp) $\ell_0$, we
 are led to $\ell_0\sim b$. Notice that $T$ is a sample-independent quantity.

 In the quantum case, a detailed numerical analysis was carried out
 for the strongly incommensurate case. For any quasi-energy $\lambda$
 a transmission
 coefficient $T(\lambda)$ was computed following the procedure outlined in
 Sec.\ \ref{sec:level2}.
 $T(\lambda)$ is a fluctuating, i.e. sample-dependent, quantity. The
 main facts about the statistics of this quantity that emerge from numerical
 computations are:

(1)  $\langle T(\lambda) \rangle_{q_0}$ (the average of
$T(\lambda)$ over $q_0$)
is but weakly dependent on $\lambda$, and is
therefore close to the total transmission (the average of $T(\lambda)$ over
$\lambda$), which is the quantum counterpart for the classical transmission
(\ref{trans}). It is on fact a decreasing
function of $L$ which comes close to (\ref{trans}) in a range of lengths
$\ell_0 \ll L \ll \ell$, or, equivalently,
$b\ll L \ll b^2/\hbar$, which is called
the "diffusive regime" (see Fig.1).

At larger lengths, $\langle T(\lambda) \rangle_{q_0}$ exponentially
decreases with $L$, as $\exp(-2L/\ell)$; this is the localized, or "insulator",
regime.

(2) In the diffusive regime, $Var(T(\lambda))$ (the variance of $T(\lambda)$)
is roughly independent of length \cite{BOGU1}.

Both (1) and (2) reproduce well-known aspects of electronic transmission
through quasi one dimensional disordered solids; in particular, the
scale-independence of the magnitude of transmission fluctuations observed
in the diffusive regime is the celebrated phenomenon of Universal
Conductance Fluctuations (UCF).

\subsection{The Kicked Harper Model}

The KHM is obtained by choosing
$S(p)=d\cos p, \quad V(q)=c\cos q$. It
was introduced in \cite{LEB,HOL}; the study of its
quantum diffusive properties was started by Lima and
Shepelyansky\cite{LS}, and was subsequently continued by a number of Authors
\cite{DIMACPH,GUBO,ABGRC,GEI1,GEI2} (for a review on the subject
see \cite{ABCGR}).
 It is classically chaotic and diffusive at $d,c \gg 1$, and the
statistical analysis outlined for the KRM is still valid (with appropriate
modifications).

The spectral properties of the quantum KHM at fixed quasi-position,
like those of the quantum KRM,
depend on the arithmetic nature of $\hbar$. If the latter is commensurate with
 $2\pi$, then the spectrum is absolutely continuous, and the propagation of
wavepackets is ballistic; a continuous spectral component is also present for
"not too incommensurate" values of $\hbar$. However, even for strongly
incommensurate values of $\hbar$ the spectral properties depend in a highly
nontrivial way on the value of the parameters $c,d$. Whereas exponential
localization
is still present at $d \gg c$, the motion is ballistic at $c \gg d$; a phase-
portrait, illustrating the type of motion in various region of the $(c,d)$
plane, can be found in \cite{ABGRC}. These nontrivial
properties of the KHM are connected
with its double periodicity (in $q$ and in $p$ as well)  \cite{GUBO}, a
feature which the KHM shares with the HM,
which is the object of the next section.

In particular, at $c=d$ the motion is always delocalized, and wave-packets
spread according to a law
$\langle (\Delta q)^2 \rangle \sim t^\gamma$. The exponent
$\gamma$ is
close to $1$, but it appears to depend on the value of $c=d$.
This behaviour signals the
presence of a singular-continuous spectral component \cite{Gu,GuMa} and in
fact a multifractal spectrum has been numerically found, with a Hausdorff
dimension
close to $0.5$.
\par
We have investigated  the question, how is this (possibly anomalous) diffusion
at $c=d$ reflected in the transmission properties of the scattering model. In
our numerical computations we have used the function $2 \arctan
(\epsilon \cos(\cdot))/\epsilon$ with $\epsilon<<1$ as an approximation
for  the function $\cos(\cdot)$ appearing
 in the generating functions $S,V$. The dependence of the classical
transmission on the sample length is shown by the dashed line in Fig.2 and by
the continuous one in Fig.3; as in the KRM, it follows the
law (\ref{trans}) with the appropriate
parameters.

The behaviour of the quantum
$\langle T(\lambda) \rangle_{q_0}$ depends on whether
$\lambda$ is chosen in the spectrum, or not.
In strongly incommensurate cases,  the quantum spectrum is singular, i.e.,
has zero Lebesgue measure, and  $\langle T(\lambda) \rangle_{q_0}$
is exponentially small at
large $L$ for almost all
values of the quasi-energy $\lambda$. Therefore, the total transmission, too,
is exponentially small at large $L$, at variance with the classical behaviour,
which follows the law (\ref{trans}) instead. Thus, as in the KRM, the
quasi-classical ohmic behaviour appears only in a limited range of lengths,
which is larger, the smaller $\hbar$. In Fig.3 the dependence
 of $\langle ln T(\lambda) \rangle_{q_0}$
on length is shown, for a strongly incommensurate case
$\hbar/2\pi=1/(6+\rho)$, where $\rho$ is the golden ratio,
$\rho=(\sqrt 5 +1)/2$, and for a value of $\lambda$ chosen at random.
After a
relatively small interval in which a rough agreement with the classical
behaviour is observed, the transmission decays exponentially. On comparing
Fig.3 with Fig.1 we see that the ohmic region is here narrower than for the
KRM, though the ratio $D/\hbar\ell_0$, which in the KRM defines the width
of this region, is roughly the same; moreover, fluctuations are definitely
bigger.

In order to detect the influence of the quantum diffusion associated with the
singular spectrum we have to take $\lambda$ in the spectrum. A convenient
strategy for locating the spectrum, and one commonly used in the study of
incommensurate systems, consists in approximating the actual, incommensurate
system with a periodic one. From a continuous fraction expansion
one obtains a sequence of rational approximants
$p_n/q_n$ to  $\hbar/2\pi=1/(6+\rho)$. The KHM with $\hbar/2\pi=p_n/q_n$
is then a periodic approximation to the strongly incommensurate KHM with
 $\hbar/2\pi=1/(6+\rho)$. The spectrum of the incommensurate KHM
is approximated by the spectrum of the periodic KHM; the latter consists of
bands separated by gaps, and the bands are generated as follows.
Due to periodicity in $q$, the quasi-momentum $\beta$ is also conserved,
 besides the quasi-position $\alpha$. At fixed $\beta, \alpha$ the dynamics
is described by a unitary matrix of rank $q_n$ \cite{GUBO};
as $\beta$ is varied at fixed
$\alpha$, the eigenphases of this matrix sweep the bands in the spectrum of
the evolution operator ${\hat U}_\alpha$, for the given
$\alpha$ and for the
given periodic approximation. \par

Due to absolute continuity of the spectrum, the long-time dynamics of the
periodic KHM is ballistic,
$(\Delta q)^2 \sim t^2$. Nevertheless, over fixed,
finite scales of length and time the
incommensurate dynamics will be reproduced by the periodic ones, the better,
the higher the chosen approximant. This is illustrated in Fig.4, where
the dependence on time of $(\Delta q)^2(t)$ is shown for the infinite KHM for
two rational approximants; two lines showing the $\sim t$ and the $\sim t^2$
behaviour are drawn for comparison.  $(\Delta q)^2(t)$ was computed as
the expectation value of $q^2$ over the wave function obtained at time $t$
by iterating the KHM propagator
$$
{\hat U }= e^{(ic/\hbar) \cos(\hat{q}_\alpha)} e^{(id/\hbar)\cos(\hat p)}
$$
with
$\alpha=0$
on the initial state $
\vert 0\rangle$.\par
Therefore, in order to analyze the scattering in a finite sample, we have
chosen a
suitably high approximant, and have located the spectrum by a combination of
two methods: first, by computing the bands by direct diagonalization of
matrices, as mentioned above; second, by analyzing the behaviour of the
transmission $T(\lambda)$ as a function of $\lambda$. An example of the results
obtained with this procedure is shown in Fig.5, where a scan of
$T(\lambda)$ over a range of $\lambda$ encompassing one of the bands obtained
by diagonalization is shown.\par

Statistical ensembles were generated by varying $\alpha$ in
the interval
$(0,2\pi)$.

We have then studied the dependence of
$\langle -\ln T(\lambda)\rangle_{\alpha}$
on length $L$ for
$\lambda$ in the center of a band; the result is shown in Fig.2.
The long-time ballistic propagation
(entailing a length-independent transmission) is not yet  manifest in the
inspected
range of lengths, as confirmed by the simulation of the infinite KHM. Open
circles are well fitted by the full line of equation :

\[-\langle \ln(T)\rangle_{\alpha} = -1.47 +1.14 \cdot \ln(\hbar L)\]

giving once more
a conductance inversely proportional to a power of length with
an exponent close to $1$, that could not be more precisely determined from
our data. Thus the diffusion (with an exponent also close to $1$) associated
with the
multifractal spectrum of the KHM results in an Ohm-like dependence
of transmission on $L$.

However, this quantum diffusive behaviour of transmission is quite different
from the
classical diffusive behaviour, represented by the dashed line, and must
therefore be considered as a purely quantum effect. A quasi-classical
behaviour is to be expected only in a region of very small $L$, (hardly
detectable with the parameter values used for Fig.2), like in Fig.1.

An obvious question is whether the difference between the classical and the
quantum transmission curves couldn't be simply imputed to a difference
between the classical and the quantum diffusion coefficients. Therefore we
have computed the quantum diffusion coefficient for the case of Fig.2 from
a direct simulation of the infinite KHM; the result was in fact different from
the classical coefficient
$(D_q \sim D_{cl}/3)$,
but still, on substituting this value in eqn.
(\ref{trans}), we could not reproduce the quantum curve. We have therefore to
conclude that the quantum transport associated with the multifractal spectrum
is intimately  different from classical diffusive transport.\par
Finally, we emphasize that on choosing different quasi-energies in the
spectrum different transmission curves are obtained (compare
for instance open and full circles in Fig. 2).
 Therefore, averaging  over quasi-energies
(for example over the bands) destroys the Ohmic dependence.

\subsection{The Harper model}

In order to put the above results into a broader context, we now describe some
results about transmission in the HM. This model does not belong
in the class of kicked models discussed
in Sec.\ \ref{sec:level2} and does not possess a
chaotic classical limit: still, under appropriate conditions it exhibits a
multifractal spectrum and quantum diffusion.\par
The Harper equation, also called almost Mathieu equation,
 is a 1--d tight-binding equation:
\FL
\begin{equation}
(\hat{H}_h \psi )_n = \psi_{n+1} +\psi_{n-1} + V_0\cos(Qna+\beta)
\psi_n = E\psi_n
\end{equation}
where $\psi_n$ is the wave function at site $n$, $a$ is the lattice
size, $\beta$ is a  phase, $E$ the energy and
$V_0$ the potential strength. Its diffusive properties have been
studied by a number of authors (for
exhaustive reviews see Ref.\cite{SOKOR} and
Ref.\cite{HIRAMOTO}), and we shall here mention only a few characteristics
connected with its transport properties.
Aubry and Andr\'e \cite{AUBRY} have proved
that this model presents a metal--insulator transition for $V_0 = 2$,
when $Qa/2\pi$ is a ``strong'' irrational number (duality theory).
This means that if $V_0 < 2$ the spectrum is absolutely continuous
with extended eigenstates, while for $V_0 > 2$ the spectrum is
pure point with localized eigenstates.
At the critical point $V_0 = 2$ the spectrum is singular
continuous with zero Lebesgue measure\cite{CMP}.

If $Qa/2\pi = p/q$, $p,q$ integers, the Harper model is periodic
\cite{SOKOR};  if  $Qa/2\pi$ is a "bad" irrational (a Liouville number)
Avron and Simon\cite{SIMON} have proved that the spectrum is
singular continuous
(instead of pure point) for $V_0 > 2$.

The dynamical implications of these findings were studied in
\cite{HIRAMOTO,GEISEL} by solving the
 time--dependent Schroedinger
equation:
\[
i\hbar {{ \partial {\psi}_n}\over{\partial t}} = (\hat {H}_h \psi )_n
\]
It was found that
at the critical point $V_0 = 2$
the spreading of wave packets obeys $Var(n) \sim t^\gamma$ with $\gamma$
very close to 1
which means diffusive motion. \par
The behaviour of transmission
at the critical point has been investigated by many authors.
Sokoloff \cite{SOKOLP} pointed out that
this model has, at the critical point,
for $Qa/2\pi$ a ``strong'' irrational number and $E=0$ (which belongs to
the spectrum), very unusual transmission properties, namely wide
fluctuations of the transmission coefficient  as a function of the length
(up to three order of magnitude for variations of the length of
order $1\%$).

At the same time  it was remarked \cite{SOKOR,OSTLUND}
that the transmission is an highly oscillatory function
of the phase $\beta$ too.

An average resistance
\[ R_{av} (L) = {1\over L} \sum_{i=1}^L R(i) \]
where $R(i)$ is the resistance of the sample of length $i$,
was then introduced in Ref.\cite{AZBEL} and used
by many authors, essentially as
a means for  distinguishing localized states from extended
ones\cite{HIRAMOTO,HALSEY,YOUYAN,SUTHERLAND,R6,GUPTA}.
The average resistance was found to oscillate with the
sample length
 for extended states and to increase exponentially for localized ones
but, to the best of our knowledge, there were no claims about its
behaviour as a function of length at the critical point. In Ref.\cite{R6}
a Renormalization Group analysis was used to find
a scaling relation for the average resistance
near the critical point; the average resistance was found to
diverge with the length as $2-{V_0} \to {0}^{+}$.
In a closely related system, the Fibonacci Model, Sutherland and
Kohmoto\cite{SUTHERLAND} have found
a power law growth for the resistance.

We have found that, if the average over the
phase $\beta$ is taken, an Ohmic dependence is observed.
To calculate the transmission coefficient we used the well known
transfer matrix method (see for instance Ref.\cite{SOKOLP}).
As in the previously considered models, the scattering problem is one for
waves propagating
on an infinite 1-d lattice.
Free waves have the form $\exp(ikna)$ and
the corresponding dispersion law is $E=2\cos(ka)$. The potential
$V_0$ is different from zero
only for $n=1,...,L-1$. The reflection and transmission
coefficients are then found connecting the
right-hand wave--function
$\psi_n =e^{-ikna} +r e^{ikna}$ for $n \geq L$
with the left-hand  wave--function  $\psi_n = t e^{-ikna}$
for $n < 1$. The transmission coefficient is
given by $T = \vert t\vert^2$.

We have fixed $E=0$, (which belongs to the
spectrum of the infinite HM because of symmetry),
$Q a/2\pi = \rho= (\sqrt{5}-1)/2$,
and $V_0 = 2$ (critical point).
In Fig.6 we show the average $\langle -\ln(T)\rangle$ versus
$\ln (L)$. The best linear fit is the dashed line,
 with slope $1.02 \pm 0.03$ and intercept $0.5\pm 0.4$.
Thus we have an average "ohmic" law, with
oscillations superimposed, which are associated with the rational
approximants to the golden mean. In
the same Fig.6 we also plot data related to three
different rational approximants (the
$7^{th}$, $9^{th}$ and $13^{th}$) to the golden mean.
These data follow the
continuous line up to a length $L \sim q$ (where $p/q$ is the
rational approximant); at larger lengths they saturate, i.e., transmission
becomes independent of length, as expected of
a pure periodic system.\par

As a function of $\beta$ , $T$ shows erratic fluctuations,
and the distribution of
$\ln(T)$ is well
approximated by a log--normal distribution (for $L > 1000$),
see Fig.7.
The lognormal distribution of the transmission coefficient
is a nontrivial occurrence, because the spectrum is
singular continuous and not pure point (the localized phase
is known to exhibit a lognormal transmission distribution).

Averaging over $\beta$ the logarithm of $T$ rather than $T$ itself
was quite effective in significantly reducing fluctuations.
Huge fluctuations of $\langle T \rangle_{\beta}$
are indeed observed when the sample length is equal to a multiple of
the denominator of the rational approximants to $Qa/2\pi$,
see for instance the full line on Fig. 8.
This effect is particularly evident at small length, where only a
few "resonances" are present.

On checking the stability of these results against changes of the energy
$E$ and of the modulation $Qa$, two relevant aspects emerge. On one hand,
no significant changes appear if the energy is changed inside the spectrum,
but far from the band edges,
(this was done as in section III(B)). On the other,
 a different irrational modulation
(quadratic or transcendental numbers) yields a different power law,
 as can be inferred from the Table \ref{table1}. For brevity we call such a
dependence
a "generalized Ohm's law", and the corresponding exponent an
 "ohmic exponent (OE)".

The OE in the third
column of Table I is obtained from the best linear fit of
 $\langle -\ln(T) \rangle_{\beta}$ vs
$\ln(L)$; the error in the last digit
is indicated in brackets, and the continuous fraction
expansion is given in the 2nd column. \par
The degree of irrationality
 of $Qa/2\pi$, rather than its absolute value,
appears to  determine the exponent of the algebraic decay of the
transmission. In fact two different numbers
 only differing in the first terms of their continued fraction expansion
(the golden mean and $[100,1,1,1,\ldots]$)
produce practically the same exponent.
The fact that in Table I the exponent 1 (or very close to 1) occurs only for
these two numbers raises the question, whether such an exponent is connected
to a noble irrational modulation. Different choices of  $Qa$, even in the
class of
the irrational algebraic numbers, produces a different OE.\par

The last point is crucial, because, as reported in
\cite{HIRAMOTO-A}, the dynamical exponent ruling  the spreading in
time of wave packets in the infinite Harper model is practically independent
of  $Qa$ (provided it is sufficiently irrational), and so appear to be the
scaling
properties of the spectrum \cite{TANG}.

This means that the OE is not directly
connected with the spectral properties, and must therefore be determined by
the structure of the eigenfunctions.
Scaling properties for the latter have been found\cite{R6}, which however do
not hold for generic irrational (e.g., transcendental) modulation, and cannot
therefore by themselves explain the generalized Ohm's law, which is observed
even for such numbers (bottom of Table I).

\section{Transmission Fluctuations}
\label{sec:level4}

In the previous sections we have studied the implications that a diffusive
spread of wave-packets, whether of a quasi-classical origin or due to a fractal
spectrum, has on the average transmission. In this section we shall instead
analyze the fluctuations of the transmission.

In the theory of transmission
through disordered solids, the quantum diffusive regime is characterized by
Universal Conductance Fluctuations, i.e., by transmission fluctuations of
scale-independent magnitude
\cite{IMRY,PICH1,PICH2,MELLO}
\cite{M1,M2,M3,M4,M5,M6,M7,M8,M9,M10,M11,M12,M13,LEE}.
On defining the dimensionless
conductance as $G=\chi T$, where $\chi$ is the number of scattering channels,
it was found that
\[
Var(G) \approx { {2}\over{15} }
\]
for a time-reversal invariant transmission process\cite{MELLO,IMRY,PICH2,LEE}
through a quasi-one dimensional sample of length $L$,
with $\ell_0 \ll L \ll \ell$.
Such universal fluctuations are considered a distinctive mark of a proper
quantum diffusive regime, besides the validity of Ohm's law (apart from
weak-localization corrections).\par

As mentioned in Sec.\ \ref{sec:level3}(A), the KRM exhibits
UCF\cite{BOGU,BOGU1}. In Figs.9,10  we present results
for  the other two models examined in this
paper which  do not share this property. Data for the KHM (Fig.9)
 are rather ambiguous.
In fact, even if the points are distributed along the line
$2/15$ (their average value is $0.12$) they appear at the
same time slowly decreasing with length.
This behaviour is especially clear with the HM (Fig.10),
, where
a power law dependence on length is found: $Var(T) \sim L^{-\delta}$
with $\delta = 0.38 \pm 0.04$ (in this case conductance and
transmission coefficient are the same since only the "in"
and "out" scattering channels are active).

One more feature preventing fluctuations from being length-independent
 is the presence of sharp
variations of $Var(T)$ as a function of the sample length $L$, when
$L$ is some multiple of a denominator occurring in rational approximants
of $Qa/2\pi$. This behaviour does not appear to be due to insufficient
statistics, and is shown by a dashed line in
Fig.8.
Therefore, although the range of lengths inspected in Figs.9-10
is ``ohmic'' for transmission,
transmission fluctuations are not universal, because $Var(T)$ is
length-dependent.
UCFs appear then to be typical of quasi-classical diffusion.

Finally we remark that in the Harper model the transmission
remains a non self--averaging
quantity\cite{LEE}, since
$Var(T)/\langle T \rangle^2$ diverges algebraically with $L$
 when $L\to\infty$.

\section{CONCLUSIONS}
\label{sec:level5}
The Ohmic dependence of transmission on the sample length is a typical
feature of classical one-dimensional diffusive transport, which can be
observed in strictly deterministic classical systems in the presence of
a chaotic dynamics. On the other hand, one-dimensional chaotic
diffusion is typically suppressed, in the long run,  by quantization,
so that an ohmic behaviour
can be quantally observed only on length scales which are small in comparison
to the quantum localization length. In some cases, however, localization
effects leave room for an unbounded pseudo-diffusive spread of wave packets,
connected with a fractal structure of the energy
(or quasi-energy) spectrum. On analyzing  the transmission properties
associated with such a "fractal" pseudo-diffusion, which is observed
in a model endowed with a chaotic classical limit, we have found evidence that
this kind of quantum diffusion is a purely quantum effect,
as unrelated to quasi-classical
chaotic diffusion as is quantum localization itself. On comparing the
behaviour of the Kicked Harper model, which exhibits both types of diffusion,
with the Kicked Rotator on one hand, and with the Harper model on the other,
it becomes apparent that the proper ohmic regime is still restricted to a
 range of "small" sample lengths, in which the singular continuous
nature of the spectrum is still unresolved (as it was the singular point-like
spectrum of the Kicked Rotator), and the quantum motion is essentially a
quasi-classical diffusion. Beyond that, the purely
quantum pseudo-diffusion alone is at work, leading to a substantially
different type of transmission, which is still power-law dependent on the
length, but with an exponent which depends on the arithmetic nature of the
Planck's constant, according to as yet unknown rules, and cannot therefore be
meaningfully carried to the classical limit. In addition, the
statistics of transmission fluctuations are quite different from the
quasi-classical case; in particular, their magnitude is not scale-independent,
as it should in a proper metallic regime. In the language of the theory of
quantum chaotic scattering, the crucial aspects of the Ericson regime of
strongly overlapped scattering resonances appear to be lost as soon as the
continuous (singular) nature of the spectrum comes into play.

\acknowledgments
We thank Roberto Artuso for help
in numerical computations, and Dima Shepelyansky for useful discussion.

\begin{figure}
\caption{Transmission versus the rescaled sample length (in units
of the mean free path) for the KRM. Full symbols are quantum data
for $b\tau = 10$, $b = 58$, $\hbar=1$.
Each point is obtained by averaging over an ensemble of $10^2,
10^3$ different samples. The quasi--energy $\lambda$ is fixed.
Full line is the classical theoretical prediction $(3.1)$.}
\end{figure}

\begin{figure}
\caption{Transmission versus the sample length
for the KHM and
$ c = d = 10$, $\hbar/2\pi=144/1097$.
Full and open circles are relative to two
quasi--energies in the middle of two different bands. Full line
is the best fit for open circles.
Each point is obtained by averaging over an ensemble of $10^2$
different samples. Dashed line is the
corresponding classical line from a kinetic equation like
$(3.1)$ with the classical diffusion coefficient.}
\end{figure}

\begin{figure}
\caption{Transmission versus the rescaled sample length (in units
of the mean free path) for the KHM. Full symbols are quantum data
for $c=d=10$,  $\hbar=2\pi/(6+\rho)$, $\rho$ the golden mean.
Each point is obtained by averaging over an ensemble of $500$
different samples. The quasi--energy $\lambda$ is fixed.
Full line is the classical theoretical prediction.}
\end{figure}

\begin{figure}
\caption{Energy growth in time for the KHM, for $c = d = 10$
and rational $\hbar/2\pi$ approximant
to $1/(6+\rho)$, $\rho$ being the golden mean. Full line is for
$\hbar/2\pi=55/419$, dots are for $\hbar/2\pi=144/1097$.
The quasi--position is fixed $\alpha = 0$.
For the sake of comparison $t$ and $t^2$ lines are shown
in dashed style.
}
\end{figure}

\begin{figure}
\caption{Transmission coefficient as a function of quasi--energy for
the KHM.
Here $c = d = 10 $,
$\hbar/2\pi=5/38$, $L=120$.
}
\end{figure}

\begin{figure}
\caption{Transmission versus length for the HM. Here $E=0$, $V_0=2$,
$Qa/2\pi =\rho$, $\rho$ the golden mean. Each point is
obtained by averaging over an ensemble of $10^3$
random phases $\beta$. Circles, squares and triangles are respectively
for the $7^{th}$, $9^{th}$, and the $13^{th}$ approximant to the
 golden mean.
Dashed line is the best fit line with slope $1.02(3)$.
}
\end{figure}

\begin{figure}
\caption{Distribution of the transmission coefficient for the HM,
at fixed length $L=1000$, by varying randomly $\beta$ in
$[0,\pi]$,
for $E=0$, $V_0=2$, $Qa/2\pi=\rho$.
The full line is the gaussian obtained from the standard
best fitting procedure.
}
\end{figure}

\begin{figure}
\caption{$\beta$-averaged transmission coefficient (full
line)
and its variance (dashed line) as a function
 of the length $L$ for the same data as Fig.7
As vertical lines are shown multiples of the rational approximants
to $\rho$: $13,21,34,55,...$.
Averaged values were computed by taking $5000$ random
$\beta$ in the interval $[0,\pi]$.
}
\end{figure}

\begin{figure}
\caption{$Var(G)$ versus the length for the KHM.
Open and full circles are obtained by fixing
two different quasi--energies and varying the quasi--position
over 100 samples; triangles are obtained by varying the
quasi--energy over 100 samples and fixing the quasi--position
$\alpha = 0$. Horizontal dashed line is the
theoretical value $2/15$.
}
\end{figure}

\begin{figure}
\caption{$\ln(Var(T))$ versus the logarithm of the
length for the HM.
Here $E=0$, $V_0=2$,
$Qa/2\pi =\rho$. Each point is obtained by averaging over
$10^3$ different samples.Dashed line is the best fit line
with slope $\delta=-0.38 \pm 0.04$.
}
\end{figure}

\begin{table}
\caption{Generalized Ohm's exponent for various irrational
modulations.}
\begin{tabular}{lrl}
$Qa/2\pi$& Expansion&$OE$\\
\tableline
$(\sqrt{5}-1)/2$&$[1,1,1,1,\ldots]$&$1.02(3)$\\
$(197-\sqrt{5})/19402$&$[100,1,1,1,\ldots]$&$1.00(5)$\\
$\sqrt{2}-1$&$[2,2,2,2,\ldots]$&$1.35(4)$\\
$\sqrt{3}-1$&$[1,2,1,2,\ldots]$&$1.15(8)$\\
$(\sqrt{3}-1)/2$&$[2,1,2,1,\ldots]$&$1.15(3)$\\
$(\sqrt{37}-4)/3$&$[1,2,3,1,2,3,\ldots]$&$1.28(4)$\\
$e-2$&$[1,2,1,1,4,\ldots]$&$0.75(3)$\\
$\pi-3$&$[7,15,1,292,1,\ldots]$&$0.38(3)$\\
\end{tabular}
\label{table1}
\end{table}

\end{document}